\documentclass[11pt]{article}
\usepackage{times}
\usepackage{graphicx, textgreek}

\usepackage[numbers,sort&compress]{natbib}

\begin{document}

\twocolumn[{\LARGE \textbf{Comment on ``On biological signaling'' by G. Nimtz and \\ H. Aichmann, Z. Naturforsch. 75a: 507–509, 2020\\*[0.2cm]}}
{\large Andrew D. Jackson and Thomas Heimburg*\\*[0.1cm]
	{\small The Niels Bohr Institute, University of Copenhagen, Blegdamsvej 17, 2100 Copenhagen \O, Denmark}\\*[-0.1cm]
	
{\normalsize \textbf{ABSTRACT}\hspace{0.5cm} In 2005, we proposed that the nerve pulse is an electromechanical soliton \cite{Heimburg2005c}. This concept represents a challenge to the well-known Hodgkin-Huxley model \cite{Hodgkin1952b} which is of a purely electrical nature. The soliton theory was criticized by Nimtz and Aichmann in a recent article in Zeitung f\"ur Naturforschung A \cite{Nimtz2020}. Here, we wish to comment on some statements that we regard as misinterpretations of our views. }}

\vspace{0.2cm}

\noindent\footnotesize{*corresponding author, theimbu@nbi.ku.dk }\\

\vspace{0.1cm}]

\normalsize


According to the soliton theory, the nerve signal can be described by an exponentially localized adiabatic pul-se reminiscent of sound propagating in the membranes of nerve axons. As early as 1852, Helmholtz observed that the propagation velocity of nerves is close to the propagation velocity of sound in a hydrocarbon layer \cite{Helmholtz1852}. The pulse-like shape of the soliton results from the non-linear nature of the nerve membrane compressibility close to a phase transition in the membrane \cite{Muzic2019}. During the pulse, the membrane is partially moved through a phase transition from a liquid disordered membrane state to a solid ordered state. The associated changes in the thickness of the membrane and the length of the axon, and the reversible release of the latent heat have all been found experimentally (reviewed in \cite{Heimburg2020}). As correctly noted by Nimtz and Aichmann, this situation allows for an immediate explanation of anesthesia due to the depression of the freezing point that they produce in membranes. We have discussed this previously in depth \cite{Heimburg2007c, Graesboll2014, Wang2018}.

Nimtz and Aichmann state in their article: \emph{``This process would represent a local transition to the lower entropy of the gel phase and would result in a negative entropy step and also a reconstruction of the hydrogen bonds geometry at this part of the membrane. However, in Heimburg’s model the bound water has been never considered.''} This statement is incorrect. We base our argument on the experimentally observed changes in heat capacity and the measured relation to the lateral compressibility. Since only the area and the lateral pressure are controlled during the measurement, all free variables of the membrane can change. The compressibility therefore contains changes in the membrane electrostatics and the change in head-group hydration. The fact that area is controlled does not imply that the other free variables are fixed. E.g., charge, polarization, membrane potential, hydration and protonation of the membrane may well change during compression. The same applies to the measurement of the heat capacity. Here the temperature is controlled, but other variables such as area, charge and membrane potential are free to change during the experiment. These changes are included in the measured values of the heat capacity.

During the propagation of the soliton, no negentropy is generated. The entropy does not change in any adiabatic processes, which include all propagating processes described by analytical mechanics or hydrodynamics (including both waves and solitons). The compression of the hydrocarbon chains of the membrane obviously chan-ges the configurational entropy. However, in an adiabatic compression of a membrane, this is compensated by a local heating of the membrane because
\begin{equation}
dS=\left(\frac{\partial S}{\partial A}\right)_T dA + \left(\frac{\partial S}{\partial T}\right)_A dT \equiv 0 \;.
	\label{eq:01}
\end{equation}

Since $(\partial S/\partial T) = c_A/T$, where $c_A$ is the heat capacity at constant area, we obtain

\begin{equation}
dT=-\frac{T}{c_A}\left(\frac{\partial S}{\partial A}\right)_T dA \;.
	\label{eq:02}
\end{equation}

During the propagation of a soliton, the temperature of the nerve environment changes because the latent heat of the transition is reversibly released and reabsorbed by the membrane environment (including the associated water) as expressed by the heat capacity \cite{Mosgaard2013a}. These temperature changes have been found experimentally \cite{Ritchie1985, Heimburg2020}.

Nimtz and Aichmann further state: \emph{``The medium of a membrane is not a pure elastic one, as mentioned above, the medium is based on polaritons. The ions and the molecules are coupled as polaritons. Their characteristics are measured by their dielectric properties, and estimating the size of the soliton should be taken into account.''} The authors are incorrect in stating that this has not been taken into account. We have shown in detail, how the change in a transmembrane field will change the membrane area, and vice versa \cite{Mosgaard2015a}. In a measurement of the lateral compressibility, membrane charge and polarization will change in order to minimize the free energy. Similarly, the surface chemistry will adjust to a given membrane area. Thus, the compression modulus will automatically contain electrical contributions that depend on transmembrane voltage and on surface hydration. It is not meaningful to distinguish between mechanical, electrical or chemical solitons. For this reason, adiabatic pulses typically display variations of all extensive and intensive variables. In the case of a membrane, the area compression is associated with changes in capacitance because both the area and the thickness of the membrane change in a transition by about a factor of 2 \cite{Heimburg2012, Zecchi2017}. A charged membrane would thus display variations in transmembrane voltage during the pulse. Therefore, the nerve soliton could be considered to be a piezoelectric or electromechanical pulse, even though such designation would tend to obscure the generality of the thermodynamic soliton. It is likely that the protonation of the membrane and the structure of surface water also change.

Nimtz and Aichmann also refer to E. Schr\"odinger's famous 1944 essay entitled \emph{What is life?} \cite{Schroedinger1944}. Here, Schr\"odinger states that ``How would we express in terms of the statistical theory the marvellous faculty of a living organism, by which it delays the decay into thermodynamical equilibrium (death)? We said before: `It feeds upon negative entropy', attracting, as it were, a stream of negative entropy upon itself, to compensate the entropy increase it produces by living and thus to maintain itself on a stationary and fairly low entropy level.'' Similar ideas had been stated earlier by Boltzmann in 1886: ``The general struggle for existence of animate beings therefore does not primarily consist in a struggle for matter {\ldots} or energy (which is available to every body in surplus in the form of heat), but in a struggle for entropy, which is provided by the transition of energy from the hot sun to the cold earth.'' (Translated from \cite{Boltzmann1905} p. 40). Such considerations eventually lead to the concept of dissipative structures in complex systems as proposed by Ilya Prigogine and collaborators (e.g., \cite{Nicolis1971, Prigogine1984}).

It is true that the entropy of a system in equilibrium must be lowered by the excitation of any adiabatic oscillation, wave, or pulse \cite{Heimburg2017}. However, the entropy remains constant during adiabatic propagation. This is different from the dissipative structures originating in non-equilibrium situations that Schr\"odinger and Boltzmann had in mind. A soliton is not a dissipative structure. It rather represents a special case of a classical analytical mechanics problem. And it seems that it provides a fairly good description of the nervous impulse.

\setlength{\bibsep}{0pt plus 0.3ex}
\small{

}
\end{document}